\documentclass[10pt,a4paper,preprint]{elsarticle}
\usepackage{amsmath,amssymb}
\usepackage{graphicx}
\biboptions{sort&compress}
\begin{document}
\begin{frontmatter}
\title{Evolution of nuclear structure in and around Z=50 closed shell: Generalized Seniority in Cd, Sn and Te isotopes}
\author{Bhoomika Maheshwari\corref{mycorrespondingauthor}}
\cortext[mycorrespondingauthor]{Corresponding author}
\ead{bhoomi@um.edu.my}
\author{Hasan Abu Kassim}
\author{Norhasliza Yusof}
\address{Department of Physics, Faculty of Science, University of Malaya, 50603 Kuala Lumpur, Malaysia. \\
Center of Theoretical Physics, Department of Physics, Faculty of Science, University of Malaya, 50603 Kuala Lumpur, Malaysia.}
\author{Ashok Kumar Jain}
\address{Amity Institute of Nuclear Science and Technology, Amity University UP, 201313 Noida, India.}
\begin{abstract}
We study the quadrupole moments and the B(E2; $2^+ \rightarrow 0^+$) values for the ${11/2}^-$ states and the first $2^+$ states, respectively, by using a multi-j generalized seniority approach in the Cd ($Z=48$), Sn ($Z=50$) and Te ($Z=52$) isotopic chains. The g-factor trends have also been discussed. Although, Cd and Te isotopes represent two-proton hole and two-proton particle systems, thus involving both kind of particles (protons and neutrons) in contrast to Sn ($Z=50$) where only neutrons play a role, we find that a similar model based on neutron valence space alone is able to explain nearly all the gross features and trends. This paper represents the first attempt to test the validity of the generalized seniority scheme away from the semi-magic region and appears to be surprisingly successful. The linearly varying quadrupole moments in Cd, Sn and Te isotopes, are described by using a consistent multi-j configuration. The asymmetric double-hump behavior of B(E2) values in Cd and Te isotopes are understood in a manner identical to that of Sn isotopes by using the generalized seniority scheme for the first time. No shell quenching is supported in the calculations; hence, the neutron magic numbers, $N=50$ and $N=82$, remain robust in these isotopic chains.
\end{abstract}
\begin{keyword}
\texttt{Generalized Seniority; Q-moments; B(E2) trends; ${11/2}^-$ and $2_1^+$ states; Cd, Sn and Te isotopes}
\end{keyword}
\end{frontmatter}
%\linenumbers
%%%%%%%%%%%%%%%%%%%%%%%%%
\section{Introduction}

In-spite of the tremendous progress made in the large scale shell model calculations due to the advances in computational physics and computers, simple models continue to hold a very useful place because they reveal the physics underlying various phenomena ~\cite{bohr1975}. It is, therefore, necessary to combine the physics understanding from the simple models with the results obtained from the sophisticated calculations to gain a complete understanding of a phenomenon. The nuclear shell model ~\cite{mayer1955} is key to our understanding of nuclear structure properties not only around the magic numbers but also across a whole range of nuclei. Symmetries of the shell model related to the pairing interaction near semi-magic nuclei result in emergence of the seniority scheme which simplifies the description of the experimental data, thus supplementing our physics understanding of these nuclei~\cite{racah1943,shalit1963,casten1990,heyde1990,talmi1993,isacker2014}.

The region around $^{132}$Sn is of special interest due to the proximity of two closed shells $Z=50$ and $N=82$ where the structure can be described in terms of few particles/holes coupled to the $^{132}$Sn core. This region is also important for understanding the nuclear behavior in stellar environments. Recent experiments~\cite{yordanov2013} have highlighted a linear behavior of $Q-$moments for the ${11/2}^-$ states of the neutron-rich Cd isotopes even beyond the $h_{11/2}$ orbital. This unique simplification of the complex configurations in Cd isotopes has surprised the community~\cite{wood2013} and has eventually led to extensive studies of Cd isotopes, from understanding the moments of ${11/2}^-$ states and other single-particle states in odd-A nuclei to the B(E2) values of the first excited $2^+$ states in even-even nuclei ~\cite{vancraeyenest2013, zhao2014, takacsy2014, ilieva2014, ramayya2014, lei2015, mishev2015, wood2015, banerjee2015, lei2016, koller2016, lascar2017, schmidt2017, qin2018, yordanov2018, chaudhary2018}. This is in addition to the continuing interest in Cd isotopes since decades, both theoretical and experimental ~\cite{rosner1964, silva1964, kawase1975, fogelberg1976, pandoh1992, zamfir1995, long1995, juutinen1996, buforn2000, corminboeuf2000, fotiades2000, fotiades20001, kautzsch2000, dillmann2003, bandyopadhyay2003, kadi2003, wang2003, batchelder2005, kautzsch2005, jungclaus2006, jungclaus2007, garrett2007, stoyer2007, hoteling2007, dworschak2008, rodriguez2008, batchelder2009, caceres2009, garrett2010, chamoli2011, frauendorf2011, luo2012, hossain2012}, where they were first labeled as ‘vibrators’ and then refuted this status by supporting a weakly deformed structure. These studies highlighted the evolution of nuclear structure across $Z=50$ shell closure and tried to address the questions like shell quenching, robustness of $N=82$ shell closure, deformation, evidence of vibrations etc. 

Our recent extension of the seniority scheme to the generalized seniority scheme~\cite{maheshwari2016, maheshwari20161, jain2017, jain20171, maheshwari2017} has offered a simplified view of the complex structure of various states in semi-magic nuclei and successfully explained a number of spectroscopic properties like the energy spectra, B(EL) trends and half-lives etc. We have recently extended the usage of generalized seniority scheme to understand the magnetic moments, or the g-factor trends and termed it as ‘Generalized Seniority Schmidt Model’(GSSM), where we merge the configurations suggested by generalized seniority with the Schmidt model and obtain the g-factor trends~\cite{maheshwari2019}. The generalized seniority motivated configurations consistently explain the electromagnetic features like B(EL) trends and g-factor trends.

We study in the present paper the spectroscopic properties and behavior of even-even Cd isotopes (2-proton holes) and Te isotopes (2-proton particles) in comparison to the Sn isotopes (Z=50 closed shell) by using the generalized seniority scheme. Presence of both protons and neutrons in the valence space makes the study very interesting as isospin comes into play. We also focus upon the ${11/2}^-$ states in odd-A isotopes and try to explain their properties by using the generalized seniority approach. 

We have divided the paper into four sections. Section 2 presents brief theoretical details and expressions for the excitation energy, Q-moments and the transition probabilities. Section 3 presents a detailed analysis of the empirical systematics and the calculated results for various spectroscopic properties in Cd, Sn and Te isotopes. Section 4 concludes the work.  

\section{\label{sec:level2}Generalized Seniority scheme and related expressions}

\subsection{Seniority and Generalized Seniority}

Seniority scheme is generally credited to Racah~\cite{racah1943} but Flowers~\cite{flowers1952} also introduced it almost simultaneously. Loosely speaking, seniority ($v$) may be defined as the number of unpaired nucleons for a given state. The complete details of seniority in single-j shell may be found in de Shalit and Talmi~\cite{shalit1963} and also in Casten~\cite{casten1990} and Heyde~\cite{heyde1990}. The quasi-spin scheme, introduced by Kerman~\cite{kerman1961} and Helmers~\cite{helmers1961}, for identical nucleons in single-j scheme is easier to follow and satisfies the SU(2) algebra formed by the pair creation operator $S^+$ and pair annihilation operator $S^-$. Detailed expressions and selection rules for transitions can be found in Talmi~\cite{talmi1993}. When $n$ identical nucleons in a single-j orbital are coupled to give a total spin of $J$, the reduced matrix elements in $j^n$ configuration can be transformed to the reduced matrix elements in $j^v$ configuration. 

When $n$ identical nucleons in a multi-j space are coupled to generate a total spin of $J$, the corresponding reduced matrix elements can be calculated in the generalized seniority scheme. The concept of generalized seniority was first introduced by Arima and Ichimura~\cite{arima1966} for the multi-j degenerate orbitals. The corresponding quasi-spin algebra can be obtained by defining a generalized pair creation operator $S^+ = \sum_{j} S^+_j$, where the summation over $j$ takes care of the multi-j situation~\cite{talmi1993}. Talmi further incorporated the non-degeneracy of the multi-j orbitals by using $S^+ = \sum_{j} \alpha_j S^+_j$, where $\alpha_j$ are the mixing coefficients~\cite{talmi1971, shlomo1972}. Our recent extension of this scheme for multi-j degenerate orbitals by defining $S^+ = \sum_j {(-1)}^{l_j} S^+_j$, as proposed by Arvieu and Moszokowski~\cite{arvieu1966}, led to a new set of generalized seniority selection rules and the corresponding generalized seniority reduction formulae~\cite{maheshwari2016, maheshwari20161, jain2017, jain20171, maheshwari2017, maheshwari2019}. Here $l_j$ denotes the orbital angular momentum of the given-j orbital. The seniority in single-j changes to the generalized seniority $v$ in multi-j with an effective$-j$ defined as ${\tilde{j}}= j \otimes j^\prime....$ having a pair degeneracy of $\Omega= \sum_j \frac{2j+1}{2} = \frac {(2 {\tilde{j}}+1)} {2}$. The shared occupancy in multi-j space is akin to the quasi-particle picture. However, the number of nucleons $n=\sum_j n_j$ and the generalized seniority $v=\sum_j v_j$ remain an integer. The pair-creation operators $S^+$, $S^-$ for multi-j satisfy a quasi-spin SU$_q$(2) algebra with generalized seniority as a quantum number. 

Kota~\cite{kota2017} has recently shown that for identical nucleons, which occupy $r$ number of j-orbitals, there will be $2^{r-1}$ number of pairing SU$_q$(2) algebras. Kota~\cite{kota2017} also goes on to show that for each set of $\alpha_j$ values ($= {(-1)}^{l_j}$), there exists a corresponding symplectic algebra Sp(2$\Omega$) arising from U(2$\Omega$) $\supset$ Sp(2$\Omega$) with $\Omega= \sum_j \frac{2j+1}{2}$. This one-to-one correspondence between Sp(N) $\leftrightarrow$ SU$_q$(2) leads to special selection rules for electro-magnetic transition operators connecting $n-$nucleon states having good generalized seniority. These selection rules coincide with the selection rules, obtained by us earlier ~\cite{maheshwari2016}. 

The electro-magnetic operators $T^{EL}$ and $T^{ML}$ respectively ($L=1,2,3,....$) are one-body operators for electric and magnetic multipoles, respectively. For the choice $\alpha_{j_i} = (-1)^{l_{j_i}}$, the Sp(N) $\leftrightarrow$ SU$_q$(2) algebra combined with the parity selection rules leads to the following selection rules:\\
(i) $T^{EL}$ with even, or odd $L$ will be $T_0^1$, i.e. a tensor of rank 1 or, a vector operator. However, if all $j-$orbitals have same parity, then $T^{EL}$ with odd-L will not exist, as parity change is not possible.\\
(ii) $T^{ML}$ with even, or odd $L$ will be $T_0^0$, i.e. a tensor of rank 0 or, a scalar. However, if all $j-$orbitals have same parity, then $T^{ML}$ with even L will not exist, as parity change is not possible. 

Based on the tensorial nature (quasi-spin scalar, or vector) of the EM-operators, explicit selection rules and the behavior of B(EL) and B(ML) for even, and odd L for a chain of isotopes were presented by us~\cite{maheshwari2016}. The choice of phase $\alpha_j= (-1)^{l_j}$ by Arvieu and Moszokowski~\cite{arvieu1966} was made 'for convenience'. Kota~\cite{kota2017} recently calculated a correlation coefficient $\varsigma (\vartheta_1, \vartheta_2) = {[ || \vartheta_1 ||_n || \vartheta_2 ||_n ]}^{-1} \langle [\tilde{\vartheta_1}]^\dagger \tilde{\vartheta_2} \rangle $ which gives the average cosine of the angle between the operators $\vartheta_1$ and $\vartheta_2$. For a given realistic effective interaction Hamiltonian $H$, and a pairing Hamiltonian $ H_{p} = S^{+}  S^- $, one can use the correlation coefficient to measure the closeness of $H$ with $H_p$, for a given set of $\alpha_j$'s used in defining $S^+$. Kota~\cite{kota2017} has shown that the correlation $\varsigma$ comes out to be the highest for the choice $\alpha_{j_i} = (-1)^{l_{j_i}}$ made by Arvieu and Moszokowski~\cite{arvieu1966}. Although the largest value obtained is $0.3$ only, suggesting that the realistic effective interaction $H$ are far from the simple pairing Hamiltonian $H_p$, this seems to be sufficient for the generalized seniority to be a good quantum number for the low-lying states and some special high-spin states. 

In this paper, we invoke the generalized seniority scheme by defining the quasi-spin operators as $S^+ = \sum_j {(-1)}^{l_j} S^+_j$ ~\cite{arvieu1966}, where $S_j^+ = \sum_m {(-1)}^{j-m} a_{jm}^+ a_{j,-m}$ ~\cite{talmi1993}, as also used in our previous papers ~\cite{maheshwari2016, maheshwari20161, jain2017, jain20171, maheshwari2017}. These operators enable to define a simple pairing Hamiltonian in multi-j shell of various orbits as $H= 2 S^+ S^- $, which is known to have the energy eigen values $[2s(s+1)-\frac{1}{2} (\Omega-n)(\Omega+2-n)]$ $= \frac{1}{2} [(n-v) (2 \Omega+2-n-v)]$. Here, $s=\sum_j s_j$ is the total quasi-spin of the state having generalized seniority $v=\sum_j v_j $ arising from multi-j $\tilde{j}=j \otimes j' \otimes....$ configuration, with the corresponding pair degeneracy of $\Omega= \sum_j \frac{2j+1}{2}=\frac{2\tilde{j}+1}{2}$.

\subsection{Excitation energy}

The most prominent signatures of good seniority states show up in the behavior of the excitation energy, Electromagnetic transition rates like B(EL) and B(ML) values, $Q-$moments and magnetic moments, or g-factor values. The excitation energies of good generalized seniority states are expected to have a valence particle number independent behavior, similar to the good seniority states arising from single-j shell. It is rather easy to show this by extending the proof for the single-j seniority scheme by defining a multi-j effective configuration as $\tilde{j}$. For a two-body odd-tensor interaction $V_{ik}$, we can define the two-body matrix elements for $0^+$ (fully-coupled) state in multi-j situation as $V_0= \langle {\tilde{j}}^2 J=0 |V_{ik} | {\tilde{j}}^2 J=0 \rangle $. Then the matrix element from a ${\tilde{j}}^n$ configuration to the ${\tilde{j}}^v$ configuration can be written as:
\begin{eqnarray}
\langle {\tilde{j}}^n v \alpha J | \sum_{i<k}^n V_{ik} | {\tilde{j}}^n v \alpha' J \rangle = \langle {\tilde{j}}^v v \alpha J |  \sum_{i<k}^n V_{ik} | {\tilde{j}}^v v \alpha' J \rangle  + \frac{n-v}{2} V_0 \delta_{\alpha,\alpha'}
\end{eqnarray}

where $\alpha$ represents the additional quantum number to distinguish the states having same generalized seniority $v$ and spin $J$ values. The first term in the above equation becomes $0$ if $v=0$ or $1$. Therefore, the multi-j two-body matrix elements for the ground states of even-even nuclei ($v = 0$), and even-odd/ odd-even nuclei ($v = 1$) can simply be followed as $\frac{n}{2}V_0$, and $(\frac{n-1}{2})V_0$, respectively. This tells that the ground state energies depend on the number of pair of particles coupled to $J=0$ state, where total particle number $n=\sum_j n_j$. Therefore, the energy difference between the generalized seniority $v=2$, $J \ne 0$ and $v=0$, $J=0$ (ground) states, in even-even nuclei, can be obtained as
\begin{eqnarray}
E({\tilde{j}}^n, v=2, J)- E({\tilde{j}}^n, v=0, J=0) &=& \langle {\tilde{j}}^2 J | V_{ik} | {\tilde{j}}^2 J \rangle- V_0 = \text{constant}
\end{eqnarray}
Thus, the energy difference remains independent of the valence particle number for a given multi-j configuration. For example, the first excited $2^+$ states in Sn isotopes are observed at nearly constant energy throughout the chain with a kink near the middle of the shell, see Fig.~\ref{fig:energy_evena}. This behavior throughout the valence space consisting of five orbitals, can not be explained by using a single-j scheme. The energy gap changes from $\approx 1.2$ MeV (before the middle) to $\approx 1.1$ MeV (after the middle) with a transitional kink at $\approx 1.3$ MeV around the middle, which may be understood in terms of change in effective $\tilde{j}$ before ($\tilde{j}=19/2$) and after ($\tilde{j}=23/2$) the middle of the shell. The kink may be related to the $N=64$ gap in the active valence space. However, the generalized seniority remains the same as $v=2$ for the full chain. 

This can further be generalized to odd-A nuclei having generalized seniority changing transitions, where $E({\tilde{j}}^n, v=3, J)- E({\tilde{j}}^n, v=1, J)$ will be particle number independent. One must be careful about choosing total $J$ value arising from the given multi-j configuration ${\tilde{j}}= j \otimes j^\prime...$ to maintain the underlying physics picture, as the $v=1$ states from such multi-j configuration can have the maximum $J=$ maximum $(j, j^\prime...)$. For example, the $v=3$, ${27/2}^-$ isomers show a nearly particle number independent variation with respect to the $v=1$, ${11/2}^-$ states, both arising from ${\tilde{j}}= h_{11/2} \otimes d_{3/2} \otimes s_{1/2}$ in neutron-rich Sn isotopes~\cite{jain2017}. 

The energy difference between the generalized seniority conserving states, i.e. the same generalized seniority states ($\Delta v=0$) in even-even nuclei can also be shown to be constant in a similar way. For example, ${10}^+$ isomers decay to the lower lying $8^+$ states by E2 transitions with nearly constant gamma energy for $N>64$ even-even Sn isotopes ~\cite{jain2017}. This result can further be generalized for odd-A nuclei having generalized seniority conserving transitions. For example, the $v=3$, ${27/2}^-$ isomers decay to the lower lying $v=3$, ${23/2}^-$ states with nearly constant gamma energies for $N>64$ odd-A Sn isotopes ~\cite{jain2017}. 

\subsection{The Electromagnetic transition rates and Moments} 

The selection rules and the expressions for the electric and magnetic multipole transitions in generalized seniority scheme, and their observed consequences have been discussed in our earlier papers~\cite{maheshwari2016, maheshwari20161, jain2017, jain20171, maheshwari2017, maheshwari2019}. Firm experimental evidence has been presented for the parabolic behavior for both the odd tensor as well as even tensor transitions. We recall these developments by reproducing the following expressions for quadrupole operators as:\\
(a)	For generalized seniority conserving ($\Delta v=0$) transitions 
\begin{eqnarray}
\langle {\tilde{j}}^n v l J ||\sum_i r_i^2 Y^{2}(\theta_i,\phi_i)|| {\tilde{j}}^n v l J \rangle = \Bigg[ \frac{\Omega-n}{\Omega-v} \Bigg] \langle {\tilde{j}}^v v l J ||\sum_i r_i^2 Y^{2}(\theta_i,\phi_i)|| {\tilde{j}}^v v l J \rangle
\end{eqnarray}
(b)	For generalized seniority changing ($\Delta v=2$) transitions
\begin{eqnarray}
\langle {\tilde{j}}^n v l J || \sum_i r_i^2 Y^{2}(\theta_i,\phi_i)|| {\tilde{j}}^n v\pm 2 l J \rangle  = 
  \Bigg[ \sqrt{\frac{(n-v+2)(2\Omega+2-n-v)}{4(\Omega+1-v)}} \Bigg]\nonumber \\ \langle {\tilde{j}}^v v l J ||\sum_i r_i^2 Y^{2}(\theta_i,\phi_i)|| {\tilde{j}}^v v\pm 2 l J \rangle 
\end{eqnarray}

The reduced matrix element in Eq.(3) can directly be related to the quadrupole moments $ Q= \langle {\tilde{j}}^n J ||\hat{Q}|| {\tilde{j}}^n J \rangle= \langle {\tilde{j}}^n J ||\sum_i r_i^2 Y^{2}|| {\tilde{j}}^n J \rangle $ with the following conclusions:\\
\noindent(i.) The Q-values depend on the pair degeneracy ($\Omega$), particle number ($n$) and the generalized seniority ($v$) as per the square bracket shown in Eq.(3). The $Q-$ moment values follow a linear relationship with $n$. The $Q-$ values change from negative to positive on filling up the given multi-j shell with a zero value in the middle of the shell due to $\frac{\Omega-n}{\Omega-v}$ term. This is in direct contrast to the $Q-$moment generated by collective deformation which is expected to be the largest in the middle of the shell. \\
\noindent(ii.) The dependence of the $\sqrt{B(E2)}$ with particle number $n$ in Eq.(4) for the generalized seniority changing transitions is different than the case of $Q-$ moments. The $\sqrt{B(E2)}$ values for $\Delta v=2$ transitions exhibit a flat trend throughout the multi-j shell, decreasing to zero at both the shell boundaries. A nearly spherical structure is supported at both the ends for the given multi-j shell. The corresponding B(E2) values in the case of seniority changing transitions ($\Delta v=2$) can be obtained as follows: 
\begin{eqnarray}
B(E2)=\frac{1}{2J_i+1}| \langle {\tilde{j}}^n v l J_f || \sum_i r_i^2 Y^{2}(\theta_i,\phi_i) || {\tilde{j}}^n v\pm2 l' J_i \rangle |^2
\end{eqnarray}
The involved reduced matrix elements can similarly be obtained by using Eq.(4) between initial $J_i$ and final $J_f$ states with respective parities of $l$ and $l'$ corresponding to the $\Delta v=2$ transitions. 

For completeness, the involved reduced matrix elements of ${\tilde{j}}^v$ configuration in Eq.(3) can further be written in terms of fractional parentage coefficients, $3j-$, $6j-$coefficients and radial matrix elements as 
\begin{eqnarray}
\langle {\tilde{j}}^v v l J ||\sum_i r_i^2 Y^{2}(\theta_i,\phi_i)|| {\tilde{j}}^v v l J \rangle = v \sum_{v_1,J_1} [ {\tilde{j}}^{v-1} (v_1 J_1) {\tilde{j}} J \mid \} {\tilde{j}}^v v J ]^2 (-1)^{J_1+{\tilde{j}}+J+2} \nonumber \\ (2J+1)  \begin{Bmatrix} {\tilde{j}} & J & J_1 \\ J & {\tilde{j}} & 2 \\ \end{Bmatrix} ({\tilde{j}} || r^2 Y^{2} || {\tilde{j}})
\end{eqnarray}
where the $[ {\tilde{j}}^{v-1} (v_1 J_1) {\tilde{j}} J \mid \} {\tilde{j}}^v v J ]$ denotes the fractional parentage coefficients from ${\tilde{j}}^{v-1}$ to ${\tilde{j}}^v$ configuration. The second last and last terms represent the $6j-$coefficient and the single-particle value of the given electric quadrupole operator in a given multi-j ${\tilde{j}}= j \otimes j^\prime....$. The later can be related to the involved radial matrix elements arising in multi-j ${\tilde{j}}$ configuration as follows:
\begin{eqnarray}
({\tilde{j}} || r^2 Y^{2} || {\tilde{j}})= (-1)^{{\tilde{j}}-\frac{1}{2}} (2 {\tilde{j}}+1) \sqrt{\frac{5}{4 \pi}} \begin{pmatrix}
{\tilde{j}} & 2 & {\tilde{j}} \\
-\frac{1}{2} & 0 & \frac{1}{2} \\
\end{pmatrix} <r^2>
\end{eqnarray}
where the second last term denotes the $3j-$symbol corresponding to spherical harmonics $Y^2$ in multi-j ${\tilde{j}}$ configuration and the last term $<r^2>$ represents the radial integral. 

\section{\label{sec:level4}Results and Discussion}

In this section, we discuss the generalized seniority calculated results for the first excited $2^+$ states and ${11/2}^-$ states in even-A and odd-A Cd, Sn and Te isotopes. The pair degeneracies $\Omega=9, 10$ and $12$ correspond to the configurations \{$ h_{11/2} \otimes d_{3/2} \otimes s_{1/2}$\}, \{$g_{7/2} \otimes d_{5/2} \otimes d_{3/2} \otimes s_{1/2}$\} and \{$d_{5/2} \otimes h_{11/2} \otimes d_{3/2} \otimes s_{1/2}$\}, respectively, in the following discussion.

\subsection{Particle number independent energy variation in Cd, Sn and Te isotopes}

\begin{figure}[!ht]
\includegraphics[width=12cm,height=11cm]{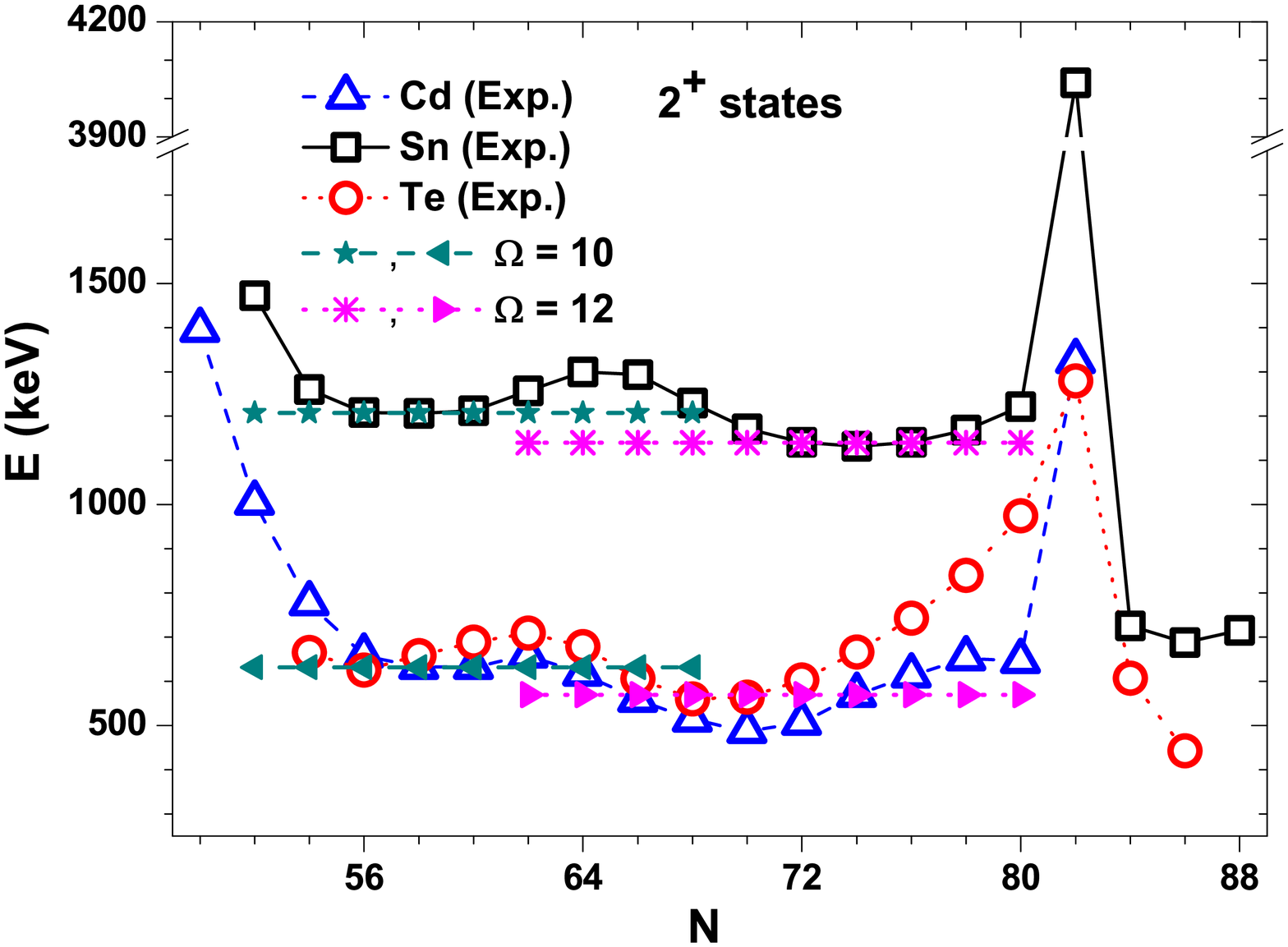}% Here is how to import EPS art
\caption{\label{fig:energy_evena}(Color online) A comparison of the experimental~\cite{ensdf} and calculated excitation energy variation of the first excited $2^+$ states in Cd, Sn and Te isotopes.}
\end{figure}

Figure~\ref{fig:energy_evena} exhibits the experimental~\cite{ensdf} energy variation of the first excited $2^+$ states with the neutron number in Cd, Sn and Te isotopes. One may note a nearly constant energy trend in all the three isotopic chains throughout $N=52-80$ with a sudden jump at $N=82$, the neutron-closed shell. The peak is quite large for the doubly magic ($Z=50$ and $N=82$) $^{132}$Sn in comparison to the other $N=82$ nuclei, $^{130}$Cd and $^{134}$Te and is yet to be fully understood. The nearly constant energies for these $2^+$ states on going from $N=52$ to $N=80$ strongly support the goodness of the generalized seniority in all the three isotopic chains. Here, the ground state has $v=0$ whereas the $2^+$ state has $v=2$. A small glitch in the energy variation around middle ($N=64$) can be noticed, which hints towards a subshell gap in the single-particle energies of the respective neutron orbitals. The active neutron orbitals in the $N=50-82$ valence space are $g_{7/2}$, $d_{5/2}$, $d_{3/2}$, $h_{11/2}$ and $s_{1/2}$, respectively. Out of which, $g_{7/2}$ and $d_{5/2}$ lie lower in energy than the remaining $d_{3/2}$, $h_{11/2}$ and $s_{1/2}$. As soon as the neutrons start to occupy this valence space, the higher probability is to occupy $g_{7/2}$ and $d_{5/2}$; however, once these two orbitals freeze out around $N=64$, the dominance of $h_{11/2}$ can be observed. So, this small change around $N=64$ is related to the change in filling of the orbitals for the first excited $2^+$ states in Cd, Sn and Te isotopes. These energy values in Cd and Te isotopes are consistently lower (nearly half) in comparison to the energies in Sn isotopes (with Z=50 closed shell), due to moving away from the closed shell. However, the generalized seniority remains constant as $v=2$ leading to a nearly particle number independent energy variation of the first excited $2^+$ states in full chain of isotopes, as shown in Fig.~\ref{fig:energy_evena}. 

We also show in Fig.~\ref{fig:energy_evena}, the calculated energy variation of the first excited $2^+$ states by using the multi-j configurations corresponding to $\Omega =10$ and $12$. The generalized seniority scheme predicts a particle number independent energy variation. The calculations have been done by using Eq.(2) and fitting the difference $E({\tilde{j}}^n, v=2, J)- E({\tilde{j}}^n, v=0, J=0)$ from one of the experimental data for the Sn isotopes, before and after the middle corresponding to $\Omega =10$ and $12$, respectively. If $V_0$ is assumed to be zero (as a normalization) for the $0^+$ ground states in even-even nuclei, then excitation energy for $2^+$ states, $E(J=2, v=2)$ may simply be related to the $\langle {\tilde{j}}^2 J | V_{ik} | {\tilde{j}}^2 J \rangle$. These calculated results as well as experimental values for Sn isotopes are far from Cd and Te isotopes, where additional contributions of two proton holes/particles are necessary. However, a constant proton contribution can take care of this nearly particle number independent energy variation in Cd and Te isotopes also, as shown in Fig.~\ref{fig:energy_evena}. This constant and additional proton contribution have been obtained by fitting $E({\tilde{j}}^n, v=2, J)- E({\tilde{j}}^n, v=0, J=0)$ from one of the experimental data of Cd isotopes for both $\Omega =10$ and $12$ multi-j configurations. Interestingly, the $2^+$ states in both Cd and Te isotopes exhibit nearly similar energies, which may be related to the proton holes/particles symmetry in the $g-$orbitals. We have, therefore, shown the results for the Cd isotopes only in Fig.~\ref{fig:energy_evena}, which seem to be in agreement with the Te isotopes up to $N=74$. Gross features of good generalized seniority are maintained in these Cd and Te isotopes. 

\begin{figure}[!ht]
\includegraphics[width=12cm,height=11cm]{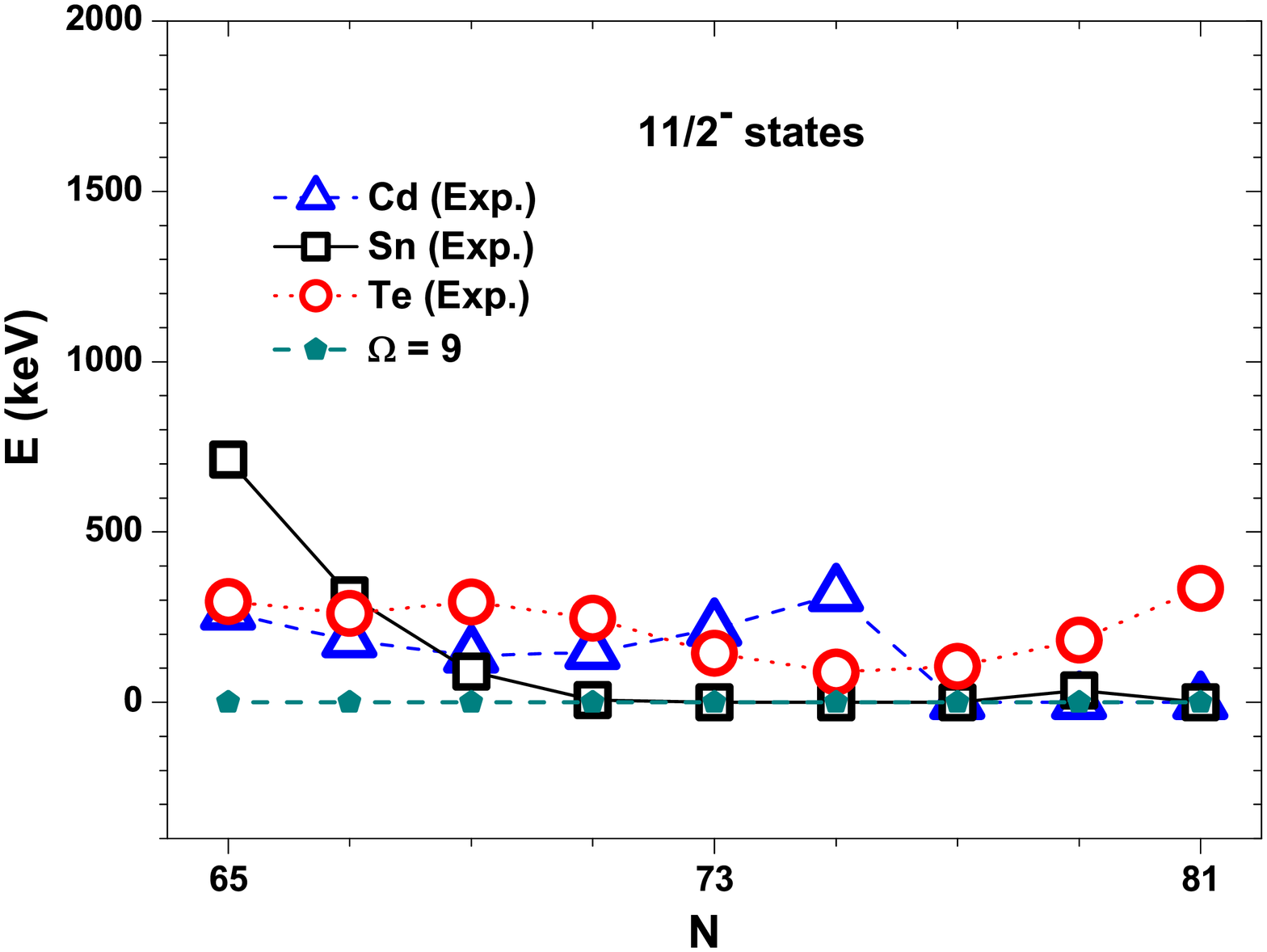}% Here is how to import EPS art
\caption{\label{fig:energy_odda}(Color online) A comparison of the experimental~\cite{ensdf} and calculated excitation energy variation of the first excited ${11/2}^-$ states in Cd, Sn and Te isotopes. The experimental value for $^{127}$Cd is shown at zero energy, by assuming X as zero in the adopted data set~\cite{ensdf}.}
\end{figure}

Furthermore, we have also shown in Fig.~\ref{fig:energy_odda}, the experimental and calculated energy variation of ${11/2}^-$ states in Cd, Sn and Te isotopes. If $V_0$ is assumed to be zero then these states must behave as the ground states of odd-A nuclei on a similar footing as the ground states of even-A nuclei. So, the calculated states are shown as $v=1$, ground states for odd-A nuclei from $\Omega=9$ configuration with $N=64$ core. A nearly particle number independent behavior is  observed in experimental data for these ${11/2}^-$ states from $N=65$ to $81$, as shown in Fig.~\ref{fig:energy_odda}.

\begin{figure}[!ht]
\includegraphics[width=12cm,height=11cm]{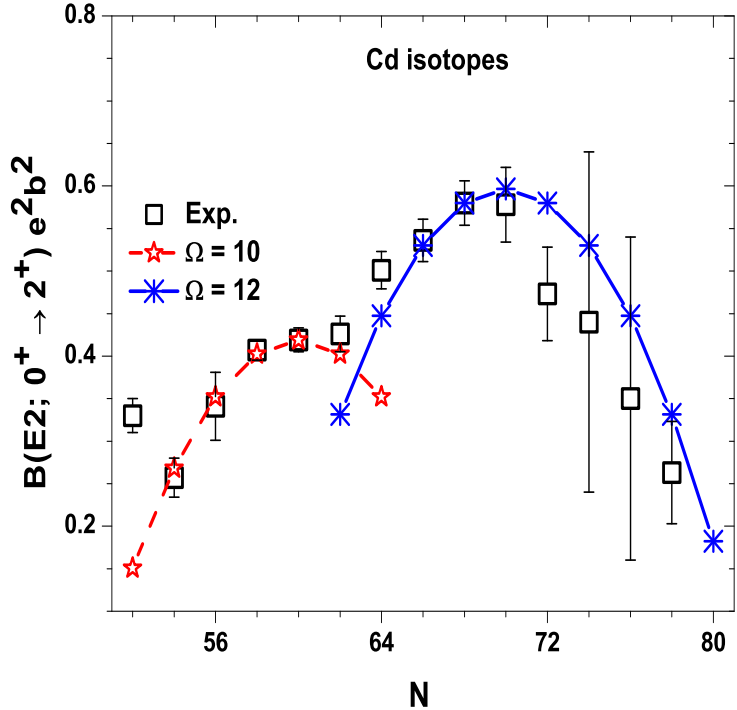}% Here is how to import EPS art
\caption{\label{fig:be2_cd}(Color online) Comparison of the experimental~\cite{pritychenko2016} and generalized seniority calculated B(E2) trends for the first excited $2^+$ states in Cd isotopes. The asymmetry in the overall trend has now been explained by the filling of different orbitals before and after the middle, resulting in a dip around middle. The chosen set of multi-j configuration in the generalized seniority calculations have been shown as $\Omega=10$ (before the middle) and $\Omega=12$ (after the middle), corresponding to $g_{7/2} \otimes d_{5/2} \otimes d_{3/2} \otimes s_{1/2}$, and $d_{5/2} \otimes h_{11/2} \otimes d_{3/2} \otimes s_{1/2}$, respectively.}
\end{figure}

\begin{figure}[!ht]
\includegraphics[width=12cm,height=11cm]{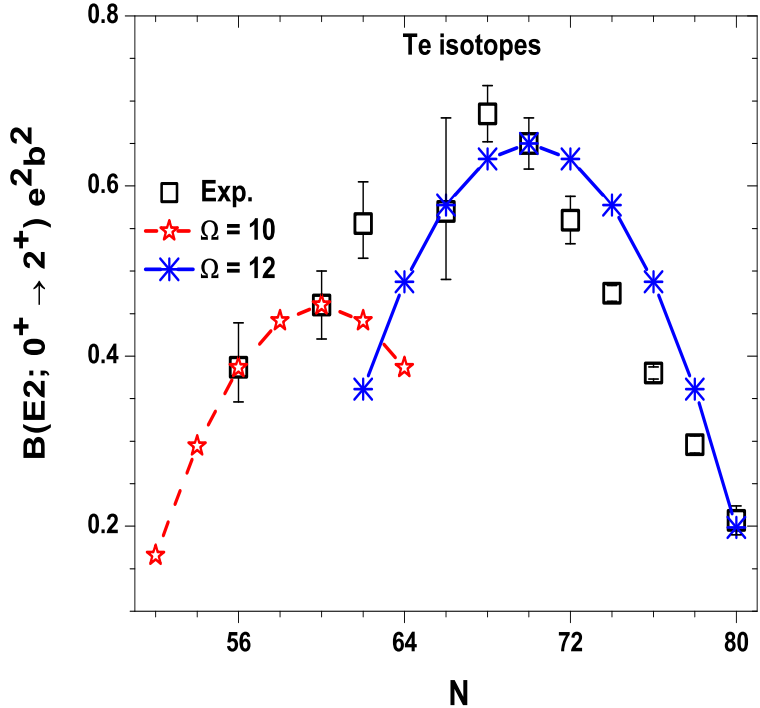}% Here is how to import EPS art
\caption{\label{fig:be2_te}(Color online) Same as Fig.~\ref{fig:be2_cd}, but for Te isotopes.}
\end{figure}

%%%%%%%%%%%%%%%%%%%%%%%%%%%%%%%%%%%%%%%%%%%%%%%%%%%%%%%%%%%%%%%%%%%%%%%%%%
\begin{table}[!htb]
\caption{\label{tab:table1} The experimental data for B(E2; $0^+ \rightarrow 2^+$) $e^2 b^2$ in even-even Cd and Te isotopes, taken from~\cite{pritychenko2016}, and the $Q$ $(b)$ data in odd-A Cd, Sn and Te isotopes, taken from~\cite{stone2014}. The uncertainties are shown in the parentheses. $Q-$ moment values for $^{113,115}$Sn are assumed with a negative sign due to systematics. }
\begin{tabular}{c c c c c c c}
\hline
\multicolumn{3}{c}{B(E2) ($e^2 b^2$)~\cite{pritychenko2016}} & \multicolumn{4}{c}{$Q$ ($b$)~\cite{stone2014}} \\
\hline
N   &  Cd &  Te & N & Cd & Sn & Te \\
\hline
&&&&&&\\
52 & 0.33(2) & & 59 & -0.94(10) & & \\
54 & 0.257(23) & & 61 & -0.92(9) &  & \\
56 & 0.341(40) & 0.387(+52-41) & 63 & -0.75(3) & -0.41(4)  & \\
   &    &    &    &   & (sign assumed) & \\
58 & 0.407(12) & & 65 & -0.61(3) & -0.38(6) & \\
  &   &     &    & & (sign assumed) & \\
60 & 0.419(14) & 0.46(4) & 67 & -0.48(2) & -0.42(5) & \\
62 & 0.426(21) & 0.556(+49-41) & 69 & -0.320(13) & & \\
64 & 0.501(22) & & 71 & -0.135(6) & -0.14(3) &  \\
66 & 0.536(25) & 0.57(+11-8) & 73 & 0.009(6) & 0.03(4) & 0.0(2)  \\
68 & 0.580(26) & 0.685(33) & 75 & 0.135(7) & 0.2(2) & 0.17(12) \\
70 & 0.578(44) & 0.650(30) & 77 & 0.269(13) & 0.32(14) & 0.40(3) \\
72 & 0.473(55) & 0.560(28) & 79 & 0.34(2) & -0.2(2) & 0.25(14) \\
74 & 0.44(20) & 0.4738(93) & 81 & 0.57(3) & 0.0(2) & 0.28(14) \\
76 & 0.35(19) & 0.3800(71) & & & & \\
78 & 0.263(60) & 0.296(10) & & & & \\
80 & & 0.207(17) & & & & \\
\hline
\end{tabular}
\end{table}
%%%%%%%%%%%%%%%%%%%%%%%%%%%%%%%%%%%%%%%%%%%%%%%%%%%%%%%%%%%%%%%%%%%%%%%%%%%%%%%%%%%%%%%%%%%%%%

\subsection{Asymmetric double-hump behavior of B(E2) values}

We could resolve the existence of two asymmetric B(E2) parabolas for the first excited $2^+$ states in even-even Sn isotopes by using the generalized seniority scheme~\cite{maheshwari20161}. In the present paper, we extend the same approach to study the B(E2) values for the transitions, from ground $0^+$ states to the first $2^+$ states, in Cd and Te isotopes. Recent B(E2) evaluation~\cite{pritychenko2016} has been used to obtain the experimental systematic trends of the B(E2) values, as listed in Table I for even-even Cd and Te isotopes. We note a nearly identical existence of two asymmetric B(E2) parabolas for the Cd and Te chain of isotopes as shown in Figs.~\ref{fig:be2_cd} and ~\ref{fig:be2_te}, respectively. A dip around the middle in the experimental B(E2) values is visible for both the Cd and Te isotopes, similar to the Sn isotopes. 

The generalized seniority calculations for the even-even Cd and Te isotopes use the multi-j configuration corresponding to $\Omega=10$ (before the middle) and $\Omega=12$ (after the middle), respectively, as assumed in the case of Sn isotopes in our earlier paper ~\cite{maheshwari20161}. The active set of orbitals is mainly dominated by $g_{7/2}$ and $d_{5/2}$ orbitals before the middle, while the $h_{11/2}$ orbital dominates after the middle. The first excited $2^+$ states have been taken as the generalized seniority $v=2$ states for calculations. The calculated trends depend on the square of the coefficients in Eq.(4), since the $0^+$ to $2^+$ transitions are generalized seniority changing $\Delta v=2$ transitions. To take care of the other structural effects, we fit one of the experimental data and restrict the values of radial integrals and involved $3j-$ and $6j-$ coeffcients as a constant, which should be the case for an interaction conserving the generalized seniority. 

It is interesting to note that the generalized seniority calculated values explain the overall trends of the experimental data in both Cd and Te isotopic chains, see Figs.~\ref{fig:be2_cd} and ~\ref{fig:be2_te}, respectively. Asymmetry in the inverted parabola before and after the middle has again been attributed to the difference in filling the two sets of orbitals. The dominance of $g_{7/2}$ orbital gets shifted to the $h_{11/2}$ orbital near the middle of the shell resulting in a dip. However, the generalized seniority remains constant at $v=2$ leading to the particle number independent energy variation for the $2^+$ states throughout the full chain. The generalized seniority, hence, governs the electromagnetic properties not only in Sn isotopes but also in the Cd and Te isotopes, which are not semi-magic nuclei. One may note that the calculations only consider the active orbitals of $N=50-82$ valence space. No signs of shell quenching have, therefore, been witnessed for these first excited $2^+$ states in all the three Cd, Sn and Te isotopes. We may note that the influence of two proton holes/ particles cannot be ignored completely; however, the overall trend is mainly governed by changing the neutron number in total E2 transition matrix elements. 

\subsection{Quadrupole moments of the ${11/2}^-$ states}

\begin{figure}[!ht]
\includegraphics[width=12cm,height=11cm]{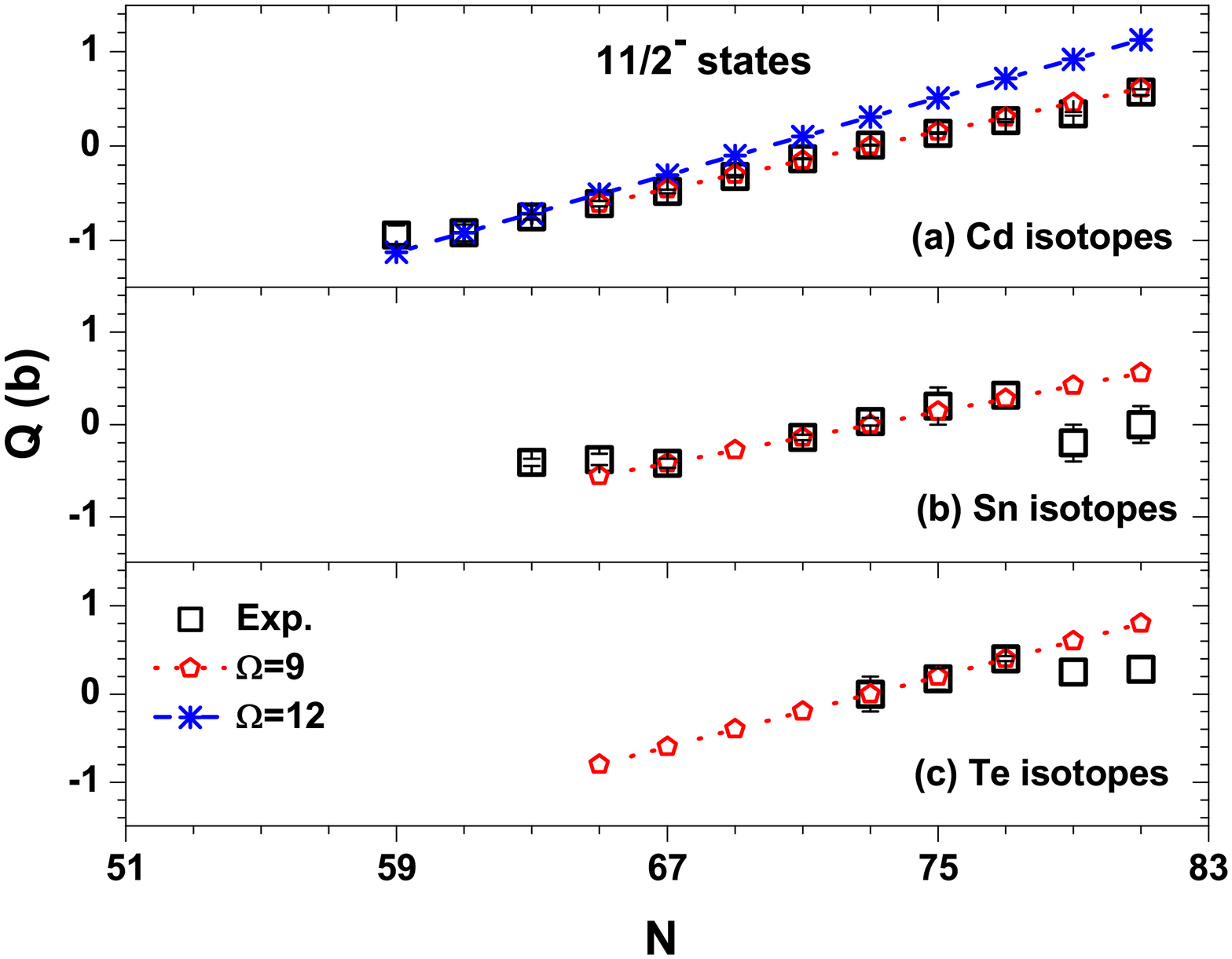}% Here is how to import EPS art
\caption{\label{fig:qmoment}(Color online) Quadrupole moment variation for the ${11/2}^-$ states in odd-A Cd, Sn and Te isotopes. All the experimental data have been taken from~\cite{stone2014}. The generalized seniority calculations by using $\Omega=9$ corresponding to the $ h_{11/2} \otimes d_{3/2} \otimes s_{1/2}$ configuration explain the experimental data for the ${11/2}^-$ states in Cd, Sn and Te isotopes, particularly after $N=64$, where the $g_{7/2}$ and $d_{5/2}$ orbitals get filled.}
\end{figure}

The $Q-$moment is usually taken as a measure of the deviation of nuclear shape from sphericity. The closed shell configurations are expected to have a nearly zero $Q-$value, while a single particle outside the closed shell may result in a negative $Q-$value. Same is true for a single hole just below the closed shell configuration which may result in a positive $Q-$value. We apply the generalized seniority scheme to study the $Q-$moments of ${11/2}^-$ states in Cd, Sn and Te isotopes. Figure~\ref{fig:qmoment} (a), (b) and (c) exhibit both the experimental and theoretical variation of $Q-$moments for the ${11/2}^-$ states, respectively, in Cd, Sn and Te isotopes. All the experimental data have been taken from ~\cite{stone2014}, as listed in Table I. The first value (from the table in ~\cite{stone2014}) in the case of multiple measurements for a given state has been adopted by us. Most of such multiple measurements for a given state are quite close to each other. The units are written as barns $(b)$, since the electric charge unit $e$ is subsumed in the definition of $Q$ by Stone~\cite{stone2014}. Such ${11/2}^-$ states are dominated by the unique-parity $h_{11/2}$ orbital in the neutron 50-82 valence space. A linear increasing trend is visible in the experimental data for all the cases, particularly after $N=64$ (a sub-shell closure after filling $g_{7/2}$ and $d_{5/2}$ orbitals). The Q-moment changes from negative to positive with a nearly zero value at $^{123}$Sn ($N=73$), $^{121}$Cd ($N=73$) and $^{125}$Te ($N=73$) in the three isotopic chains. Yordanov et al.~\cite{yordanov2013} have noted the linear behavior in the Cd isotopes beyond $h_{11/2}$ orbital. We find a similar behavior for the Sn and Te isotopes as shown in Fig.~\ref{fig:qmoment}. 

The calculations have been done by assuming these ${11/2}^-$ states as the generalized seniority $v=1$ states arising from the multi-j configuration corresponding to the pair degeneracy $\Omega=9$, as also used in our recent paper ~\cite{maheshwari2019}. Eq.(3) has been used to calculate the $Q-$moments for these generalized seniority $v=1$ and $\Omega=9$, ${11/2}^-$ states by fitting one of the experimental data. The $g_{7/2}$ and $d_{5/2}$ orbitals are assumed to be fully occupied till $N=64$, since these two orbitals lie lower in energy and get active as soon as neutrons begin to fill the $N=50-82$ valence space. The remaining three orbitals lie a bit higher in energy and start to dominate the resulting wave functions for neutron–rich isotopes. Hence, we obtain a linear trend as per the $ \frac{\Omega-n}{\Omega-v}$ coefficient for the full chains after $N=64$, as shown in Fig.~\ref{fig:qmoment}. The calculated trends are in line with the experimental data for all the three Cd, Sn and Te isotopic chains. The calculated trends also support the zero Q value at $N=73$ in all the three isotopic chains. Also, a similar range of Q values is observed with changing neutron number (from $N=65$ to $81$) for all the three isotopic chains, supporting similar structure of these states on going from Cd (2 proton-holes) to Sn (the proton closed-shell) to Te (2 proton-particles). The multi-j description in terms of generalized seniority scheme is necessary; occupancy of $h_{11/2}$ alone cannot explain the experimental data for the full chain.

For completeness, we hereby describe the present case of ${11/2}^-$, $v=1$ states in terms of the single-particle $Q-$value as expressed in Eq.(7), as the other factors will simply be equal to $1$ in Eq.(6). In this way, we can get the information of involved radial integrals for the ${11/2}^-$ state. For this, we have used the Eq.(3) and fitted the experimental Q-moment value at $^{117}$Sn ($N=67$ and $n=3$; after freezing $g_{7/2}$ and $d_{5/2}$) to find out the radial integral value for the ${11/2}^-$ state in Sn isotopes. This comes out to be $42.04$ ${fm}^2$, which can be treated as a constant for the ${11/2}^-$ states throughout the active valence space corresponding to $\Omega=9$. Similarly, one can find the radial integral value for ${11/2}^-$ states in Cd and Te isotopes, which come out to be $45.78$ and $60.05$ ${fm}^2$, by fitting the experimental value at $^{113}$Cd and $^{129}$Te, respectively. If the charge of the effective ${\tilde{j}}= h_{11/2} \otimes d_{3/2} \otimes s_{1/2}$ neutron is equal to the charge of a free neutron, we can also deduce, ${<r^2>}^{1/2}= 6.77$, $6.48$ and $7.75$ $fm$ for Cd, Sn and Te isotopes, respectively. We have also checked the radial integrals for pure-j $h_{11/2}$ orbital in all the three isotopic chains and found that the radial integral values change significantly resulting in a positive Q value, and most importantly, one can not explain the full trend.      

We note that the data for $^{129,131}$Sn ($N=79, 81$) isotopes, and for $^{131,133}$Te ($N=79, 81$) isotopes deviate from the expected trend. This is consistent with the previously noted limitations for the B(E2) explanation of ${10}^+$ and ${27/2}^-$ Sn-isomers after $N=77$~\cite{maheshwari2016, jain20171}. It may be understood in terms of limiting vacancy available in $h_{11/2}$ for these neutron-rich nuclei. So, the $d_{3/2}$ and $s_{1/2}$ orbitals can significantly alter the resulting values. It may be overcome by using a realistic picture of non-degenerate orbitals in the generalized seniority scheme. 

We have also examined the $Q-$values in lighter Cd isotopes (before $N=64$) with the generalized seniority $v=1$ and $\Omega=12$ configuration and presented a comparison in Fig.~\ref{fig:qmoment}(a). Interestingly, the calculated values explain the data in $^{109,111}$Cd but deviate in $^{113}$Cd as approaching $N=65$, and finally go off from the measured $Q-$values for heavier Cd isotopes. This comparison further validates the $N=64$ sub-shell closure where $g_{7/2}$ and $d_{5/2}$ orbitals freeze out; however, the possibility of mixing $d_{5/2}$ orbital in the resulting wave function cannot be ruled out for lighter ($N<64$) Cd isotopes. In this way, Q- moments behave as a sensitive probe for understanding the dominant role of orbitals and underlying configurations. It is important to note that the same multi-j configuration has been used to explain the trend of g-factors for these ${11/2}^-$ states in Sn isotopes very recently~\cite{maheshwari2019}, and to describe the origin of high-spin isomers like ${10}^+$ and ${27/2}^-$ in earlier works ~\cite{maheshwari2016, jain2017}. The generalized seniority scheme, hence, explains the spectroscopic features in the Cd, Sn and Te isotopes in a simple way.

\subsection{g-factor trends}

\begin{figure}[!ht]
\includegraphics[width=12cm,height=11cm]{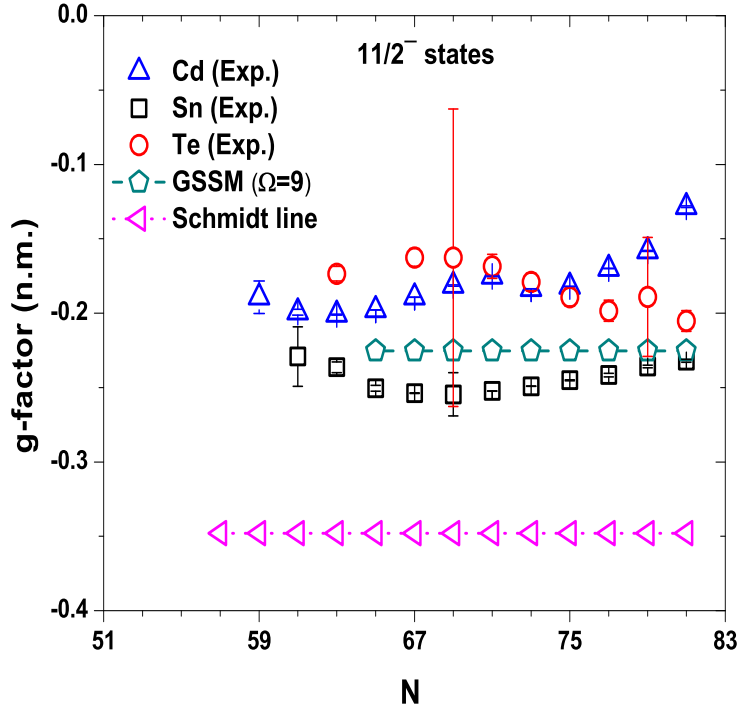}% Here is how to import EPS art
\caption{\label{fig:gfactorodda}(Color online) g-factor variation for the ${11/2}^-$ states in odd-A Cd, Sn and Te isotopes. All the experimental data have been taken from ~\cite{stone2014}. Schmidt line is shown for pure neutron $h_{11/2}$ orbital. GSSM result corresponds to $ h_{11/2} \otimes d_{3/2} \otimes s_{1/2}$ having $\Omega=9$.}
\end{figure}

We also examine the magnetic moments by obtaining the g-factors for the ${11/2}^-$ states in the Cd, Sn and Te isotopic chains. Figure~\ref{fig:gfactorodda} shows a comparison of the experimental and the Generalized Seniority Schmidt Model (GSSM) calculated g-factor trends for the ${11/2}^-$ states in odd-A Cd, Sn and Te isotopes. The data of magnetic moments~\cite{stone2014} have been used for evaluating the single-particle g-factors. The empirical trends for a given state are found to be nearly particle number independent, supporting the expectations from the generalized seniority scheme. We have recently shown that the g-factor variation of ${11/2}^-$ states in Sn isotopes can be understood as the generalized seniority $v=1$ state for $N>64$ isotopes arising from $\Omega=9$ (by freezing $g_{7/2}$ and $d_{5/2}$ orbitals at $N=64$) configuration by using the GSSM~\cite{maheshwari2019}.

\begin{figure}[!ht]
\includegraphics[width=12cm,height=11cm]{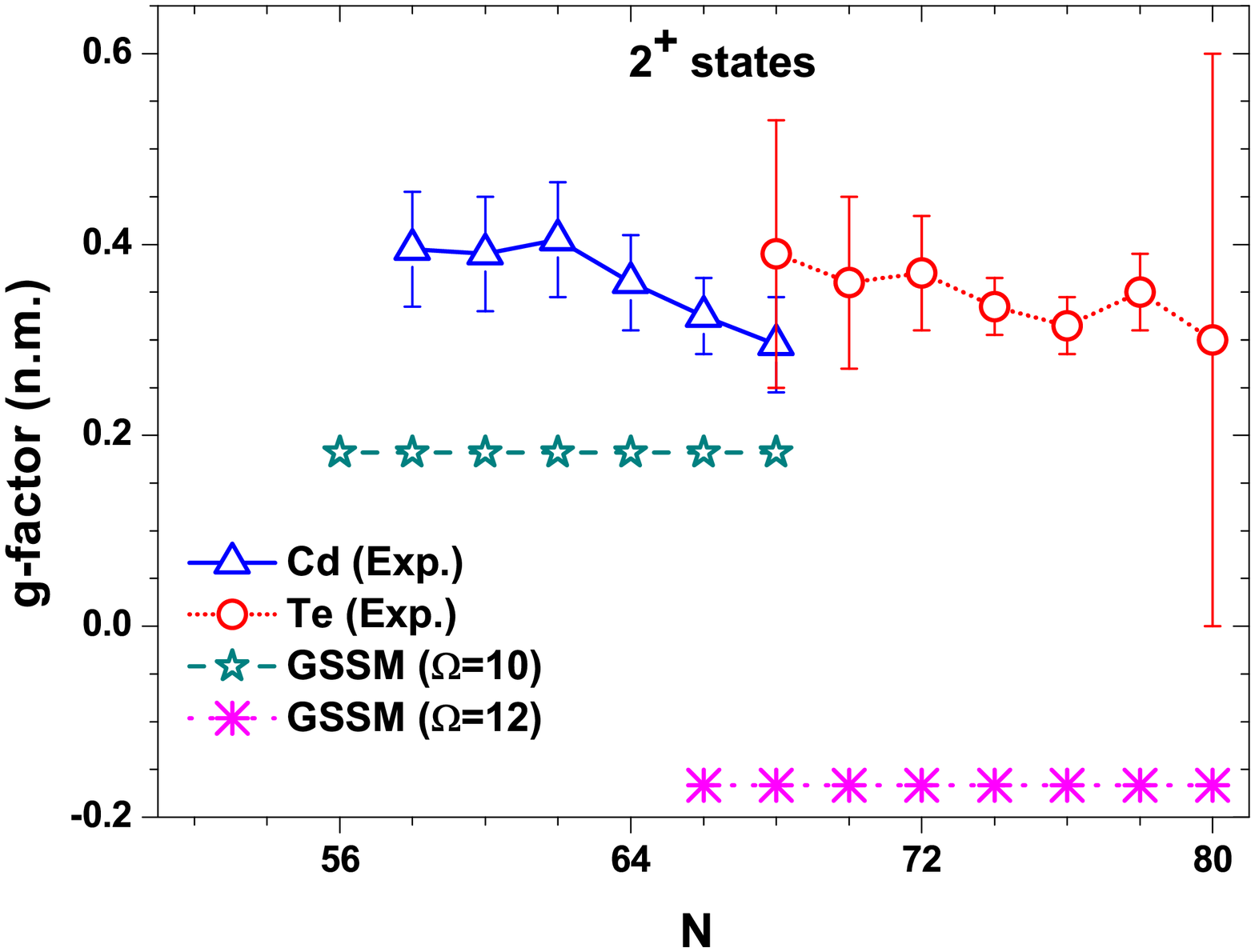}% Here is how to import EPS art
\caption{\label{fig:gfactorevena}(Color online) The empirical g-factor trend ~\cite{stone2014} for the first excited $2^+$ states in Cd and Te isotopes. The results from GSSM with $\Omega=10, 12$ from the multi-j configuration consisting of $g_{7/2} \otimes d_{5/2} \otimes d_{3/2} \otimes s_{1/2}$ and $d_{5/2} \otimes h_{11/2} \otimes d_{3/2} \otimes s_{1/2}$, respectively, have also been shown for comparison.}
\end{figure}

We now apply the similar multi-j configuration corresponding to $\Omega=9$ for the ${11/2}^-$ states in Cd ($Z=48$; two proton holes and $N>64$) and Te ($Z=52$; two proton particles and $N>64$) isotopes and find that the calculated GSSM line as defined in~\cite{maheshwari2019} lies closer to the respective experimental trends than the pure Schmidt line of $h_{11/2}$ orbital, as shown in Fig.~\ref{fig:gfactorodda}. However, the calculated GSSM line lies systematically lower than the empirical trends for both Cd and Te isotopes while it nearly follows the experimental trend in the case of Sn isotopes. Hence, contributions from proton orbitals cannot be ignored completely for Cd and Te isotopes having two proton holes and particles, respectively. The proton holes/particles are more likely to make a constant contribution throughout the complete chain. 

Similar results are also expected for the high-spin, high-seniority isomers in Cd and Te isotopes dominated by $h_{11/2}$ orbital. The consistent validity of $\Omega=9$ configuration from the two proton holes ($Z=48$) to the proton shell-closure ($Z=50$) and finally to the two proton particles ($Z=52$), is remarkable. Besides, the nearly constant behavior of g-factor of these ${11/2}^-$ states in lighter $^{111,113}$Sn, $^{107,109,111}$Cd and $^{115}$Te isotopes may be understood due to the active $h_{11/2}$ orbital along with a mixing of lower lying $g_{7/2}$ and $d_{5/2}$ orbitals. 

We also present in Fig.~\ref{fig:gfactorevena}, the g-factor trends for the first excited $2^+$ states in Cd and Te isotopes based on our recent results for the first $2^+$ states in Sn isotopes by using the GSSM~\cite{maheshwari2019}, where a positive g-factor value (with $\Omega=10$ and corresponding $\tilde{j}$) for lighter mass Sn isotopes has been found to change to a negative g-factor value (with $\Omega=12$ and corresponding $\tilde{j}$) for heavier mass Sn isotopes due to the dominance of $h_{11/2}$ neutron orbital. Fig.~\ref{fig:gfactorevena} shows the empirical g-factors for the first excited $2^+$ states in Cd and Te isotopes. Cd isotopes have two proton holes in $g_{9/2}$ while Te isotopes have two proton particles in $g_{7/2}$. The g-factor for these $2^+$ states in both the Cd and Te isotopic chains are in similar range of $0.3-0.4$ n.m. and nearly particle number independent as expected from good generalized seniority states. The pure Schmidt g-factor values for both the $g_{9/2}$ and $g_{7/2}$ protons come out to be 1.51 n.m. and 1.42 n.m. respectively, which is quite far from the empirical data. We have shown the GSSM calculated value by using $\Omega=10$ and $\Omega=12$ for comparison in Fig.~\ref{fig:gfactorevena}. It is interesting to see that all the available g-factor values in both Cd and Te isotopes are positive, which may be understood in terms of available two proton holes/particles.  

\section{\label{sec:level5}Summary and Conclusion}

We present the behavior of the quadrupole moments and B(E2) values for the ${11/2}^-$ states and first excited $2^+$ states in Cd, Sn and Te isotopes and explain them on the basis of the multi-j generalized seniority scheme. Two asymmetric B(E2) parabolas for the first $2^+$ states have been noticed in Cd (two-proton holes) and Te (two-proton particles) isotopes, and explained in terms of generalized seniority identical to the case of Sn isotopes, for the first time. The consistency of configuration on going from two-holes to two-particles is remarkable, which results in a nearly particle number independent energy variation for the first $2^+$ states in all the three isotopic chains. 

The generalized seniority scheme further explains the linear incremental trend of $Q-$moments of ${11/2}^-$ states beyond $h_{11/2}$ orbital, not only in Sn isotopes but also in Cd and Te isotopes, for the first time. New measurements are needed to confirm this situation, particularly in Te isotopes. The g-factor trends are also found to be nearly particle number independent lying close to each other, supporting the goodness of generalized seniority. Such studies, hence, provide an explanation of the structural evolution in and around Z=50 closed shell in terms of generalized seniority scheme.   No shell quenching has been witnessed. This work further suggests the need of future studies to extend the generalized seniority scheme with both type of nucleons and non-degenerate orbitals description.   

\section*{Acknowledgments}

We would like to thank Prof. B. K. Agrawal from SINP, Kolkata and Prof. V. K. B. Kota from PRL, Ahmedabad for useful suggestions. BM gratefully acknowledges the financial support in the form of a Post-Doctoral Research Fellowship from University of Malaya, Kuala Lumpur. HAK and NY acknowledge support from the Research University Grant (GP0448-2018) under University of Malaya. AKJ acknowledges the Amity Institute of Nuclear Science and Technology, Amity University UP, Noida for financial support.

%\newpage %Just because of unusual number of tables stacked at end

%\bibliographystyle{short&compress}% Produces the bibliography via BibTeX.

\end{document}